# Longitudinal spin relaxation in nitrogen-vacancy ensembles in diamond

*Authors:*
M. Mrózek[1], D. Rudnicki[1], P. Kehayias[2], A. Jarmola[2], D. Budker[2,3], and W. Gawlik[1]

*Affiliation:*
[1]Institute of Physics, Jagiellonian University, Łojasiewicza 11, 30-348 Kraków, Poland,
[2]Department of Physics, University of California, Berkeley, California 94720-7300, USA
[3]Helmholtz Institute, Johannes-Gutenberg University, Mainz, Germany

*Abstract*

We present an experimental study of the longitudinal electron-spin relaxation of ensembles of negatively charged nitrogen-vacancy (NV⁻) centers in diamond. The measurements were performed with samples having different NV⁻ concentrations and at different temperatures and magnetic fields. We found that the relaxation rate $T_1^{-1}$ increases when transition frequencies in NV⁻ centers with different orientations become degenerate and interpret this as cross-relaxation caused by dipole-dipole interaction.

## 1. INTRODUCTION

Nitrogen-vacancy (NV⁻) color centers are point defects in the diamond lattice, which consist of a substitutional-nitrogen atom adjacent to a lattice vacancy. They possess nonzero electron spin (S=1) and can be optically initialized and read out, which allows numerous applications including electric-field, magnetic-field, pressure, and temperature sensing, as well as nanoscale NMR [1-10]. Nanodiamonds containing NV⁻ centers are non-toxic, photostable, and can be easily functionalized, meaning they can be used as fluorescent markers or sensors in biological materials [11-17].

Successful applications require sound knowledge of interaction of the NV⁻ centers with environment. This motivates recent efforts to study the relaxation dynamics of the NV⁻ electron-spin polarization. Particularly, measurements of the longitudinal relaxation time $T_1$ (the decay lifetime for NVs population initialized to a ground-state magnetic sublevel) might result in development of new techniques to use NV-$T_1$ as a spin probe, for example, in biological systems [18-24].

The temperature and magnetic dependences of the longitudinal electron-spin relaxation time $T_1$ have been experimentally studied previously [25,26]. In particular, in Ref. [25] the $T_1$ relaxation was studied through the temperature dependence of the decay of the optically enhanced EPR signal, hinting at a two-photon Raman and Orbach-type processes as the main relaxation mechanisms. The results of Ref. [26] revealed temperature independent longitudinal relaxation at low temperatures where the $T_1$ relaxation is magnetic field dependent and strongly affected by the cross-relaxation between differently aligned NV⁻ centers as well as between NV⁻ centers and substitutional nitrogen impurities (known as the P1 centers).

In this paper we build upon previous work to improve our understanding of the NV⁻ $T_1$ and concentrate on detailed studies of the longitudinal relaxation in NV⁻ ensembles as a function of the strength and direction of the magnetic field. Following Ref. [26] we propose a qualitative description of the mechanisms responsible for the $T_1$ relaxation at zero magnetic field and near 595 G in terms of the dipole-dipole interactions between the neighboring centers. The dipole-dipole interactions were also considered responsible for narrow luminescence resonances detected as a function of magnetic field in Ref. [27, 28].

In Section 2 we present the experimental setup and the method of measuring the longitudinal relaxation rates; in Section 3 we describe our results of studying the magnetic-field dependence of the relaxation rates. We concentrate on the effects of varying the direction of the magnetic field and on the zero-field relaxation resonance. Section 4 is devoted to a discussion of the cross-relaxation and analysis of the zero-field resonance in terms of magnetically enhanced dipole-dipole interaction. The paper is concluded in Section 5 and supplemented by appendix which presents results of temperature measurements obtained for several samples.

## 2. EXPERIMENTAL

In our measurements we used bulk diamonds (~2×2×0.5 mm$^3$ in size) from Element 6 [Table 1]. They were manufactured by either high-pressure-high-temperature (HPHT) synthesis or chemical vapor deposition (CVD) technique. All samples were cut along the (100) crystallographic surface. All samples except S5 were irradiated at Mainz Microtron (MAMI) with a 14 MeV electron beam; sample S5 was irradiated at Argonne National Laboratory with 3 MeV. After irradiation they were annealed for two hours at the temperatures listed in Tab. 1 The values of NV$^-$ concentration in the samples were estimated by the fluorescence and absorption techniques described in Ref. [5].

| Sample | Synthesis | [N] (ppm) | [NV$^-$] (ppm) | Radiation dose (cm$^{-2}$) | Annealing (°C) |
|---|---|---|---|---|---|
| E1 | HPHT | <200 | 18 | 10$^{18}$ | 650 |
| E2 | HPHT | <200 | 20 | 10$^{18}$ | 650 |
| S5 | HPHT | <200 | 40 | 8×10$^{17}$ | 700 |
| W5 | HPHT | <200 | 2 | 1.5×10$^{17}$ | 750 |
| E8 | CVD | <1 | 0.02 | 10$^{16}$ | 650 |

Tab. 1 Details on sample preparation and characteristics.

Figure 1(a) presents the energy level structure and the relevant optical transitions that create spin polarization of the NV$^-$ ground state and allow its detection as a change of the luminescence signal. The samples were characterized by recording optically-detected magnetic resonance (ODMR) spectra, i.e. the luminescence intensity vs. the frequency of microwaves (MW) driving the transition between spin states in the ground state of NV$^-$. Measurements were performed using a confocal-microscopy setup described previously [26,29]. Figure 1(b) presents the layout of our experimental setup.

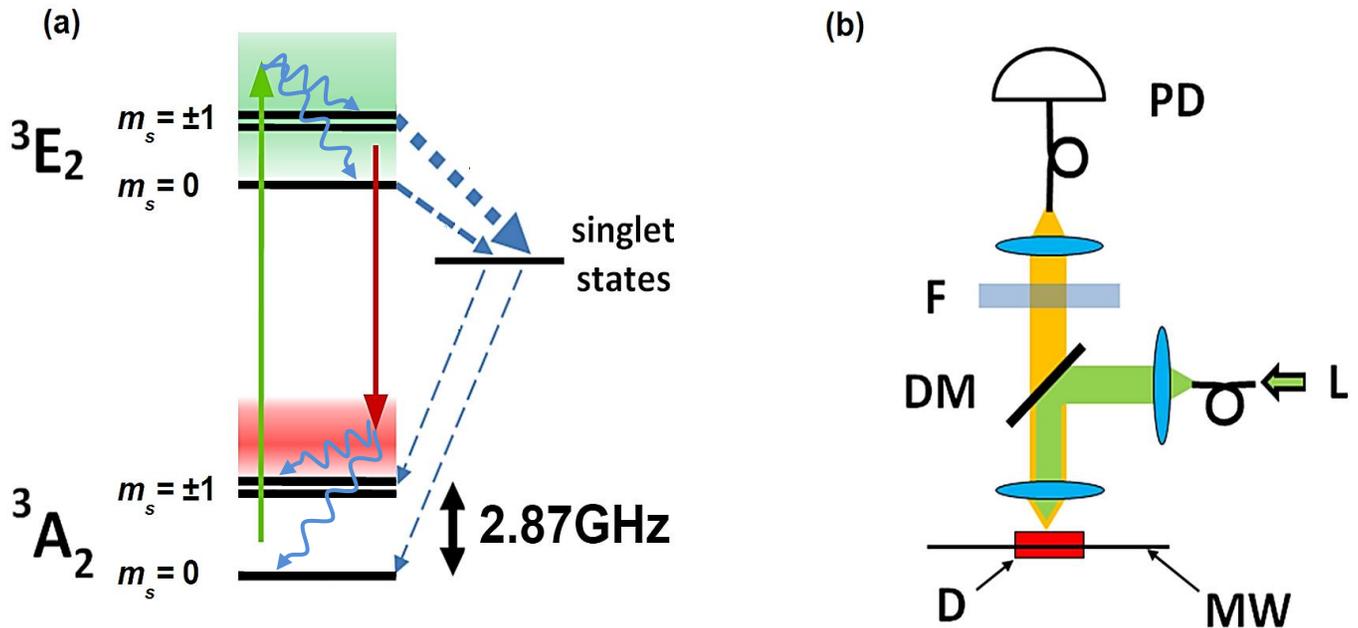

Fig. 1 (a) Energy level scheme of the NV$^-$ center in diamond. Green arrow marks the excitation by a 532 nm laser beam, the red arrow represents the optical transitions involved in the zero phonon line (637 nm), the wavy represent the nonradiative transitions, the black double arrow indicates the MW transitions, while the broken arrows symbolize the intersystem transitions (arrow thicknesses mimics the transition strengths). (b) Confocal setup, D marks the NV diamond sample, MW the microwave stripline, DM the dichroic mirror, L the green laser, F the red filter, and PD the photodiode.

The measurements were conducted in a range of temperatures between 10 and 400 K. For driving transitions between the spin states of the NV- ground state, a MW field was applied to the NV- ensemble by either a copper wire 70 um in diameter, placed on the top of diamond surface (Samples E2, S5, W5) or by a stripline printed circuit placed inside the cryostat directly under the sample (Samples E1, E8) [30].

While the ODMR data were taken with MW in CW mode, for the $T_1$ measurements, the MW power was delivered in the form of pulses, produced and controlled by a programmable pulse generator. Figure 2 shows the timing sequence of the MW and optical pulses used in the experiments to measure $T_1$. Each sequence begins with a 1 ms green-light pulse, which optically pumps NV- centers into the $m_s = 0$ state, thereby creating spin polarization of their ground $^3A_2$ state. Then a resonant MW π-pulse is applied to transfer the NV- centers into the $m_s = +1$ or $m_s = -1$ state. Following an adjustable time delay τ, another light pulse is applied and the NV- fluorescence is detected to determine the remaining spin polarization of the color centers. After such measurement, a second control sequence is applied with the π-pulse omitted, and the difference between the two results is evaluated. This common-mode rejection procedure [26] cancels most background fluorescence contributions except fluorescence from the chosen NV- subensemble. At this study, we assume that all magnetic sublevels of the ground state have the same decay rates. This assumption, however, remains yet to be verified since the decay channels between the $m_S$ states are not well identified. Measurements of single-NV- $T_1$ could be helpful to learn about individual sublevel lifetimes. Also, recent experiments on the strain coupling of the $m_S = \pm 1$ states [31] might be helpful to shed some light onto that issue by addressing these states independently on the $m_S = 0$ state.

Our study was performed with DC magnetic fields of 0 to 400 G. The field direction was controlled with a system of three pairs of Helmholtz coils or with a permanent magnet aligned along a chosen direction and mounted at an adjustable distance from the sample.

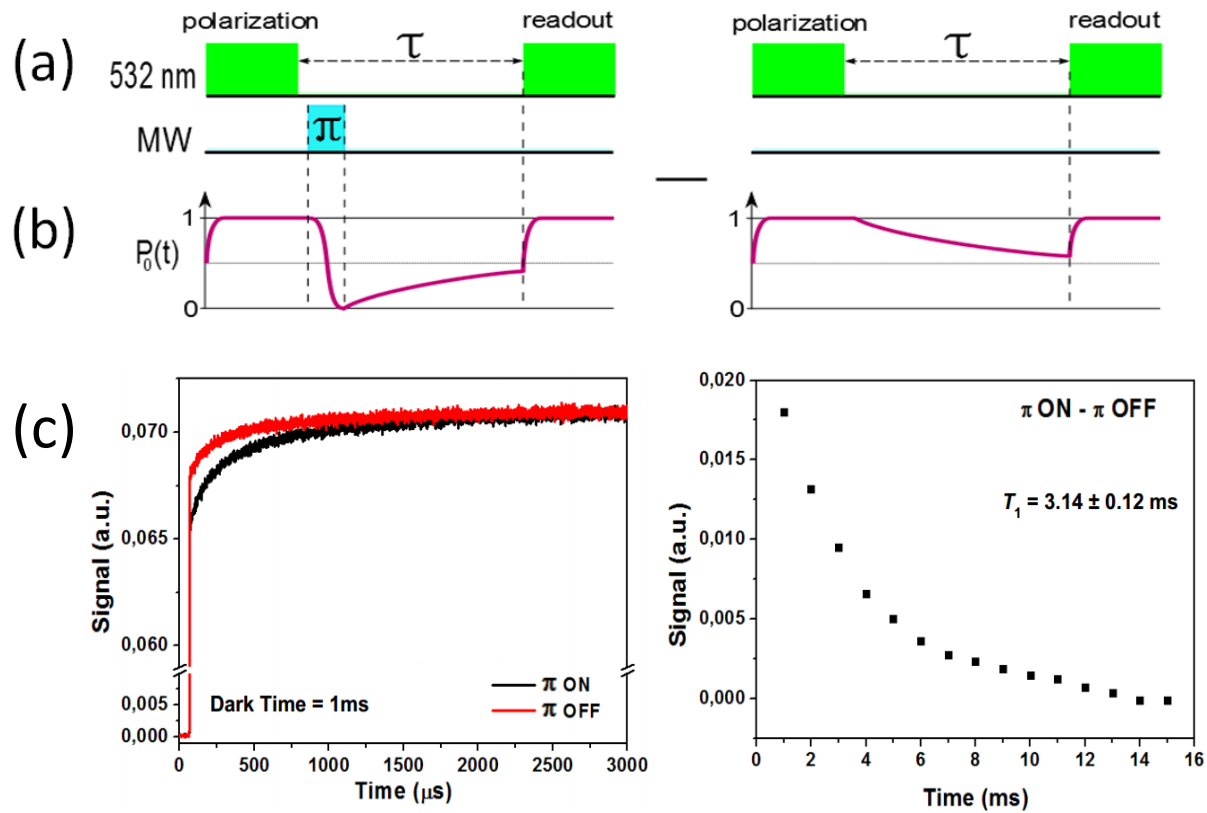

Fig. 2 (a). Optical and microwave pulse sequence used for $T_1$ measurements. The upper (green) rectangles show the optical excitation, while the lower (blue) one is the MW pulse. (b) Population $P_0(t)$ of the ground state $m_s = 0$. This state is predominantly responsible for the luminescence signal with (left panel) and without (right panel) the MW pulse. (c) Time dependence of the luminescence signals at the beginning of the readout pulse with and without the MW pulse (left panel) and their difference (right panel). Subtraction of two subsequent recordings with and without the MW pulse for different delay times τ allows determination of the longitudinal relaxation time $T_1$.

## 3. RESULTS

### 3.1. Magnetic-field dependence

The axis of NV⁻ center is parallel to one of the [111] crystallographic directions in a diamond crystal so there are four alignments of NVs that constitute four spatial subensembles. For the field directed along [111], one spatial subensemble of NVs is parallel to the field and three subensembles are at the angle of ≈109° degrees to **B**. Considering their electronic structure, the three subensembles groups are degenerate. In the field along [100] all NVs have the same angle about 55° with **B**.

Figure 3(a) depicts the ODMR spectra of an NV⁻ sample recorded with various strengths of the magnetic field applied along the [111] direction. Figure 3(b) shows transition frequencies between the magnetic sublevels of the NV⁻ color center. Solid lines correspond to the NV⁻ centers with axis parallel to the magnetic field, whereas the dashed ones are associated with the centers at ≈109° to the field.

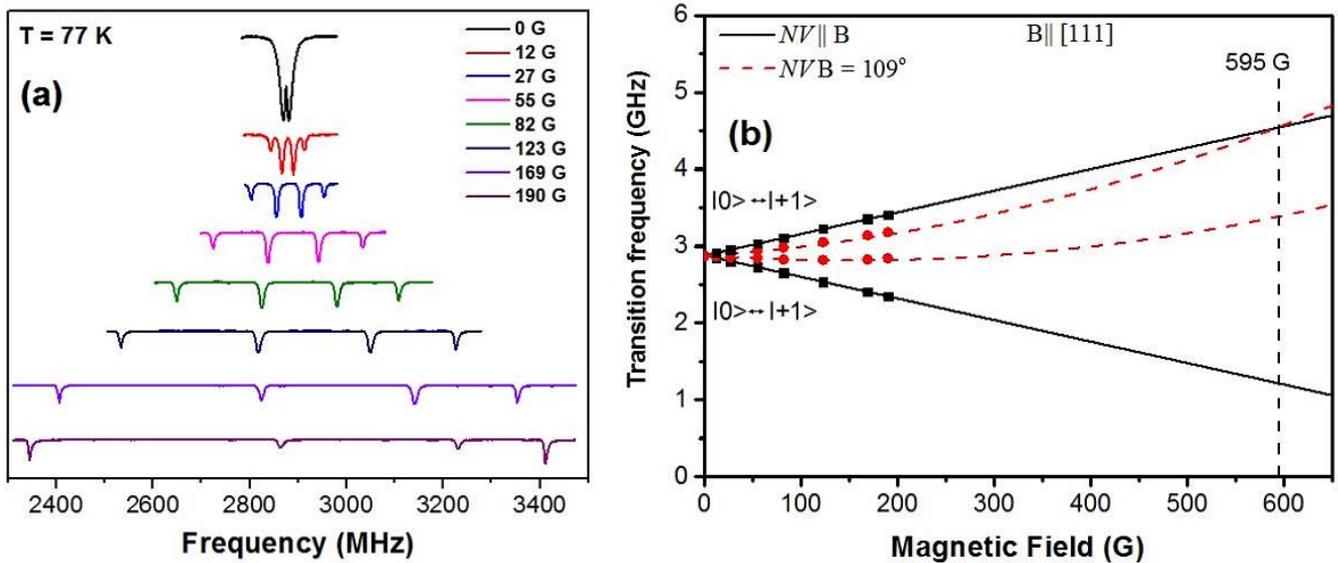

Fig. 3 (a) An example ODMR spectra for different values of the magnetic field, set in the crystallographic direction [111]; (b) transition frequencies from $m_s$=0 state of NV⁻ centers as a function of magnetic field along the [111] direction (solid lines) and at ≈109° degrees (broken lines).

At fields of 0 and ~595 G, the transition frequencies become degenerate for various orientations of the crystallographic axes, i.e. at line crossings of appropriate transitions between magnetic sublevels. As described below, the relaxation rates exhibit distinct increases at these crossing points (similar resonances have been observed near 514 G due to P1 centers [24,26]).

## a) Direction of the magnetic field

We studied the $T_1^{-1}$ rate at fixed magnetic field, B = 50 G, as a function of magnetic-field orientation relative the three crystallographic directions depicted in Figure 4.

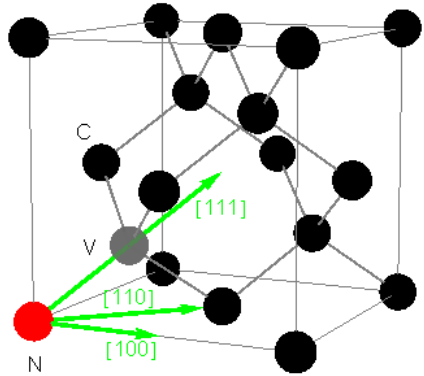

Fig. 4. Structure of the NV- diamond crystal with the orientations of the crystallographic directions

This field strength allows for a good separation of the ODMR components and is unaffected by the reduced $T_1$ near B = 0 [26]. Figure 5(a) depicts the change of the ODMR spectra upon the rotation of the B-field between the [110] and [100] directions and between the [110] and [111] directions. The particular resonances in Fig. 5 (a), well resolved for some intermediate orientation angles, are marked as I, IIa, IIb. Figure 5 (b) presents the related dependences of the measured longitudinal relaxation rates $T_1^{-1}$ on the B-field orientation at room temperature.

As shown in Fig. 5 (a), for some directions of the magnetic field between the [100] and [110] and between the [110] and [111] axes some components of the ODMR spectrum become degenerate. When such degeneracy occurs, i.e. when NVs of different orientations (other spatial subensembles) are degenerate, $T_1$ becomes shorter [Fig. 5 (b)]. Quantitatively: (i) when a fourfold-degenerate resonance I (at [100]) becomes doubly-degenerate then $T_1$ lengthens by a factor of about two [corresponding to halving the relaxation rate as seen in Fig. 5 (b)]; (ii) when the non-degenerate resonance IIb becomes triply-degenerate (at [111]), $T_1$ shortens by a factor of about four; (iii) when a non-degenerate resonance overlaps with non-degenerate resonances, IIa and IIb (at [110], $T_1$ time shortens by a factor of about two.

The effect appears analogous to the relaxation resonances that occur when electronic transitions have the same energies at B=0 and 595 G. The later resonance was studied in Ref. [26] and the former is described in more detail below and discussed in Sec. 4.

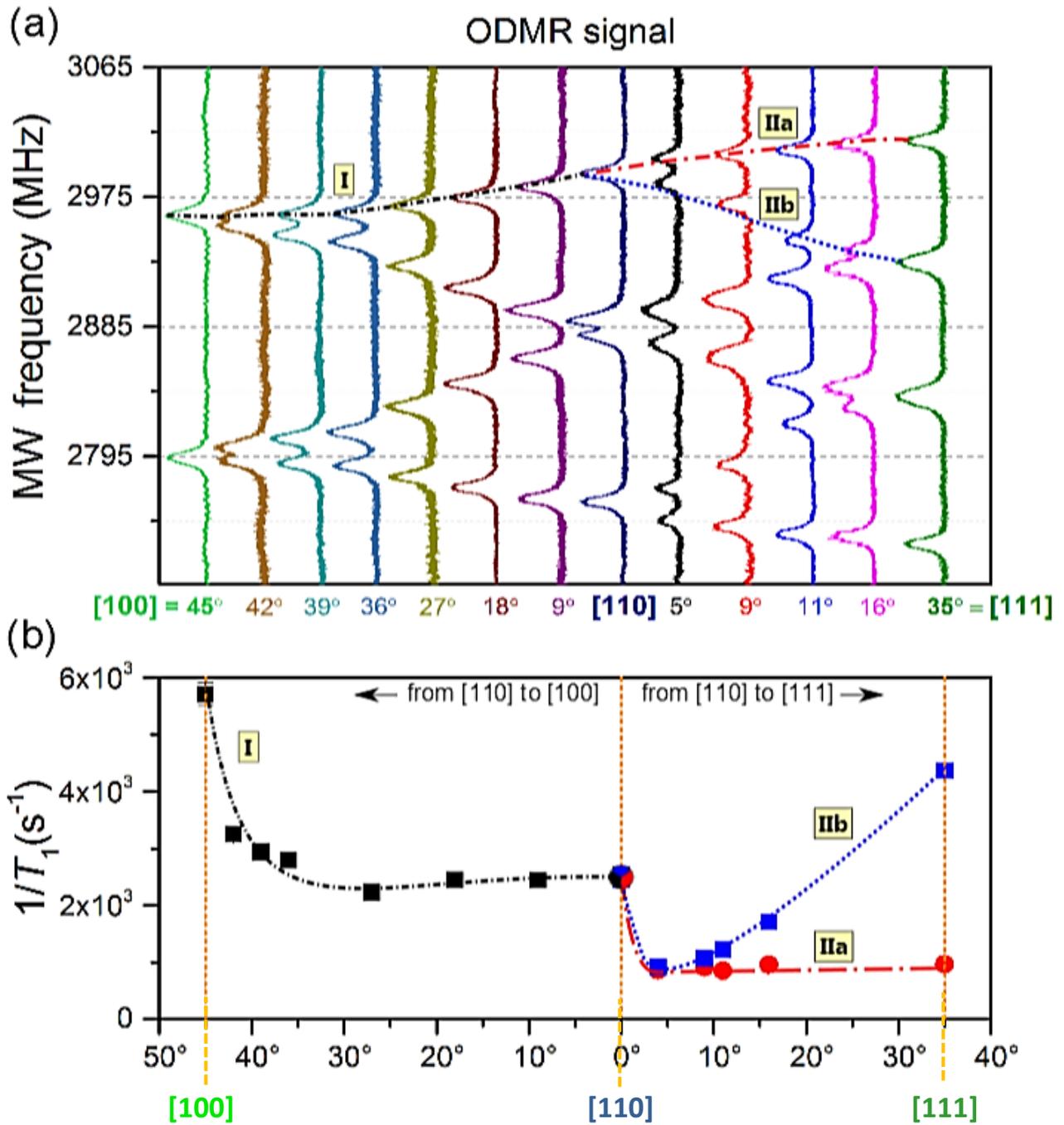

Fig. 5 (a) Sample S5 at room temperature - evolution of the ODMR spectra when the magnetic field (B=50 G) is gradually tilted between the [100] and [110] and between the [110] and [111] directions. Symbols I, IIa, IIb, label specific ODMR resonances which change their frequency as a function of the angle. (b) The related $T_1^{-1}$ relaxation rates vs. the angle between **B** and the crystallographic axes for three groups of ODMR resonances (marked by different colors). Dash/dot lines in Fig. 5 (a) and (b) are used for guiding the eye, while the vertical dotted lines indicate the crystallographic directions. For specific angles different number of spatial subensembles are degenerate, as indicated below the plots.

**b) Zero-field resonance**

With the magnetic field along the [111] direction we measured $T_1^{-1}$ for the external and internal components of the four-line ODMR spectrum associated with the NV$^-$ subensembles oriented at 0° and ≈109° degrees relative to **B**. Figure 6 (a) depicts such a spectrum obtained with low MW power (amplifier output +13 dBm). We found that the $T_1^{-1}$ rates are different for the outer and inner resonances, as depicted in Fig. 6(b) for two temperatures. The relaxation rates increase with decreasing field strength. At zero value of the magnetic field, where all ODMR components become degenerate and overlap, their relaxation rates merge to a common maximum value, i.e. exhibit the *zero-field relaxation resonance*.

The range of intermediate magnetic fields, i.e. such that the ODMR components are partly overlapping, is associated with an interesting transformation of the shape of the corresponding zero-field relaxation resonances. These shapes are different for the internal and external components of the ODMR spectrum. Particularly, the resonances corresponding to the internal ODMR lines exhibit W-like shapes resembling second derivative of a Lorentzian: the relaxation rate first drops as one approaches the resonance and then peaks around the center at B=0. We found that these unusual shapes are artefacts caused by our procedure of assigning single relaxation rates to the decay curves discussed in Sec. 2, corresponding to individual components of the ODMR spectrum. As long as the components are well resolved, we can assign them well-defined relaxation rates. In the range of weak magnetic fields, however, the ODMR components overlap and cannot be assigned with single rates. The width of the overlap of the ODMR components depends on MW power and is visibly broadened at the power level corresponding to reliable measurements of $T_1^{-1}$ (amplifier output +48 dBm). The discussed nonstandard shape of the overlapping zero-field resonances is particularly well seen in the T=77 K data in Fig. 6(c) where shading marks the region of overlapping ODMR components. Open symbols mark the relaxation rates associated with three overlapped subensembles.

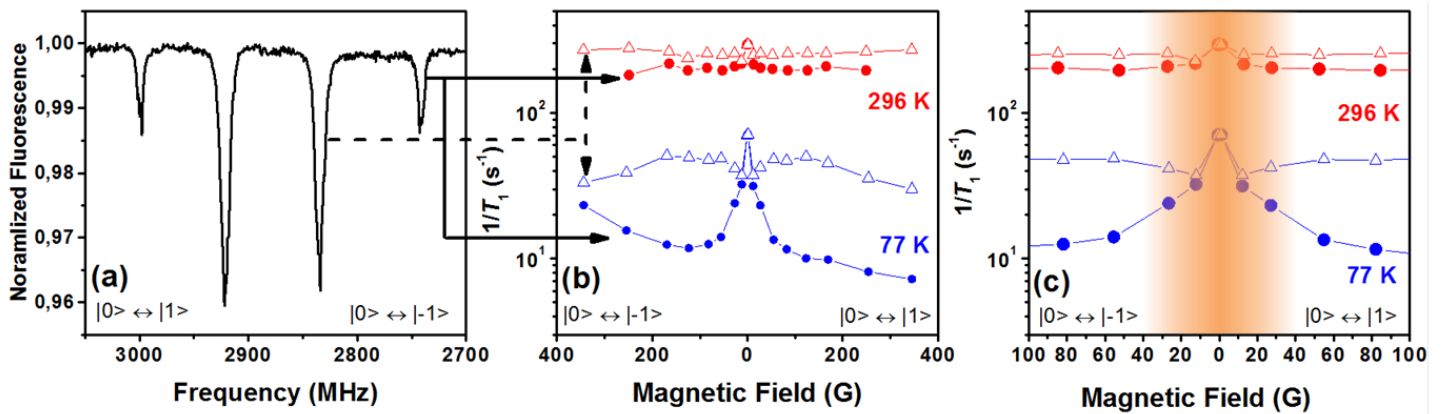

Fig. 6. (a) An example of the ODMR spectrum in a magnetic field of 44 G applied in the [111] direction (sample E2) with low MW power. At higher MW power Different relaxation rates were measured for different resonances: (b, c) relaxation rates $T_1^{-1}$ as a function of the strength of the magnetic field for the external (full dots) and internal (open triangles) components of the ODMR spectrum at 77 K (blue) and 296 K (red). (c) central region near zero field where the ODMR lines overlap (shaded area). For B=0, the external and internal ODMR components are degenerate and the points representing corresponding relaxation rates overlap.

Figure 7 presents the zero-field resonances recorded for various samples at room temperature and 77 K. Their amplitudes depend on the specific sample, in particular on their NV- concentrations. At room temperature the resonances corresponding to lower NV- concentrations are practically buried in the phonon-dominated $T_1$ [Fig.7 (a)], at 77 K, however, they are well visible [Fig.7 (b)], with the exception of the W5 and E8 samples. We attribute the differences of the amplitudes of the zero-field relaxation resonances to different NV- concentrations, which determine the strength of the dipole-dipole NV-NV couplings.

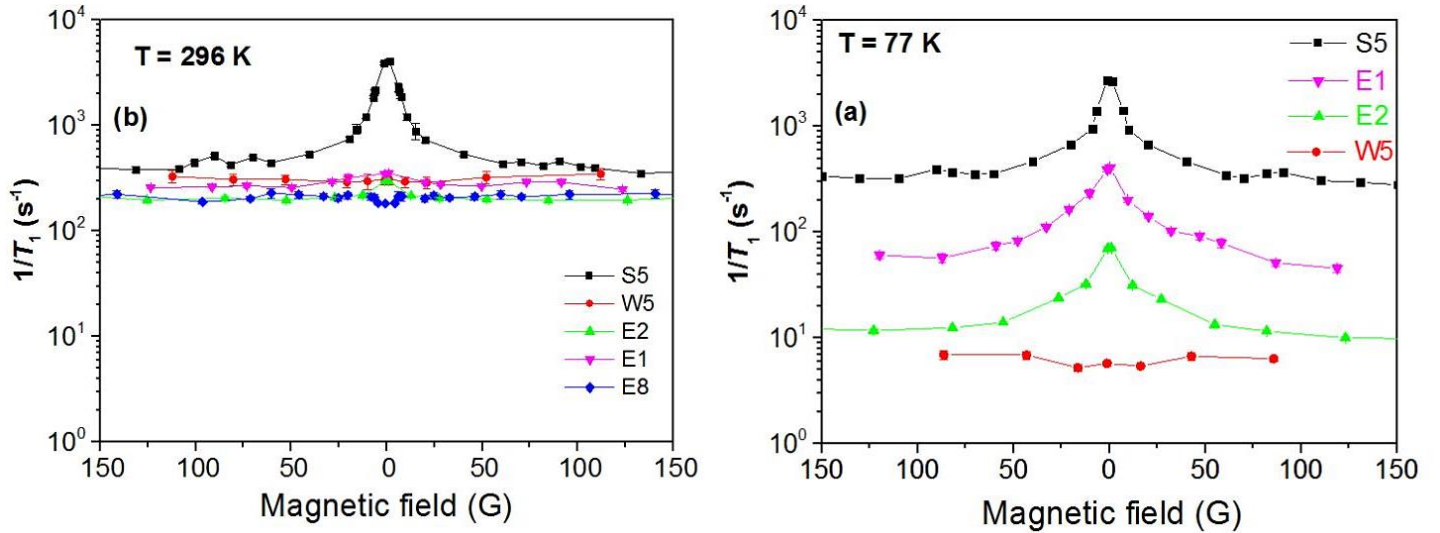

Fig. 7. The dependences of the $T_1^{-1}$ rates on the magnetic field for all investigated samples recorded at room temperature (a) and at 77 K (b).

## 4. DISCUSSION

### 4.1. Cross-relaxation

We interpret the observed relaxation resonances as being caused by cross-relaxation resulting from the dipole-dipole interaction. The strength of the dipole-dipole interaction depends on the magnetic field direction. In particular, its efficiency depends on orientation of the magnetic field relative the three, above discussed, directions in the diamond crystal: specifically, for the magnetic field along the [111] direction we can distinguish two spin sub-ensembles, one denoted as NV1, consisting of the centers aligned along the magnetic field and the other one, NV2, tilted ≈109° with respect to the magnetic field. The eigenstates of the ground state of NV1 are $|m_s=0\rangle$ and $|m_s=\pm 1\rangle$, while those of NV2, the states $|\varphi 0\rangle$, $|\varphi +\rangle$, and $|\varphi -\rangle$, are superpositions of $|m_s=0\rangle$ and $|m_s=\pm 1\rangle$. At 595 G, the frequencies of the ODMR transitions $|0\rangle \leftrightarrow |-1\rangle$ and, $|\varphi +\rangle \leftrightarrow |\varphi -\rangle$ become equal, as schematically represented in Fig. 8 by red arrows. In such case, NV2 in some state $|\psi\rangle \neq |0\rangle$ with different polarization degree, i.e. with different populations, may undergo spin-flipping processes associated with release of energy which may be absorbed by NV1 and cause its depolarization. Analogous degeneracy occurs for the transition frequencies $|0\rangle \leftrightarrow |+1\rangle$ and $|\varphi +\rangle \leftrightarrow |\varphi 0\rangle$ illustrated by blue arrows and results in an additional mixing channel and related shortening of $T_1$. Such increase of the longitudinal relaxation rate $T_1^{-1}$ due to cross-relaxation in differently aligned NVs was seen in Refs. [22,26] and interpreted as magnetic mixing caused by transverse fields.

The NV- cross-relaxation contributes to $T_1$ relaxation, and the effect is enhanced when different NV sub-ensembles are degenerate. It is thus a collective effect characteristic of ensembles, rather than a degeneracy of energies occurring within a single center. In that way it is analogous to the "line crossing" [32] rather than the "level-crossing" effect [33] known in atomic spectroscopy.

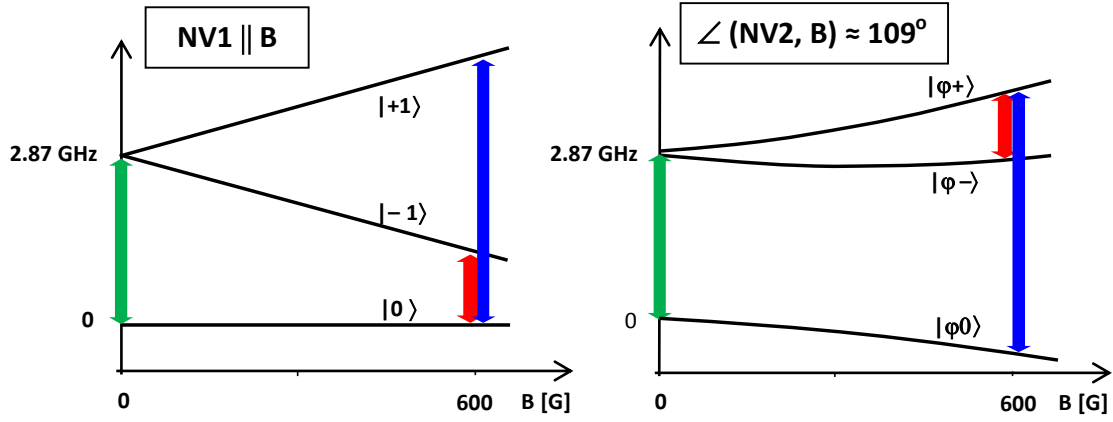

Fig. 8. Possible cross-relaxation channels for interaction between two NV⁻ centers: NV1 aligned along the magnetic field **B** and NV2 aligned at ≈109 degree relative to **B**. The solid arrows show where NV dipole-dipole $T_1$ decay is enhanced by resonance / line crossings at specific ODMR transitions which become degenerate in magnetic field of: 595 G for transitions $|0\rangle \leftrightarrow |-1\rangle$ and $|\varphi+\rangle \leftrightarrow |\varphi-\rangle$ (red), $|0\rangle \leftrightarrow |+1\rangle$ and $|\varphi+\rangle \leftrightarrow |\varphi 0\rangle$ (blue), and at zero field for both transitions (green).

The cross relaxation effect is not limited to the crossing at 595 G. Whenever a spatial subensemble of NVs (which can be seen as a single resonance in the ODMR spectra) becomes degenerate with other subensemble, the cross relaxation increases the $T_1^{-1}$ rate. We have demonstrated this effect in Fig. 5 (b) by varying the angle of the magnetic field (B=50 G) relative to the crystallographic directions. Another special case of the transition frequency degeneracy is the line crossing at zero magnetic field. Since we never achieve 100% NV⁻ polarization, the described mechanism can also play important role even in B=0 field (green arrows in Fig. 8).

## 5. CONCLUSIONS

The present study of electron-spin relaxation in NV⁻ diamond performed with various strengths and orientations of a magnetic field reveals the important role of cross-relaxation processes in NV⁻ ensembles. We find distinct resonances of the relaxation rate $T_1^{-1}$ when transition frequencies between the ground-state spin states of NV⁻ centers with different orientations become degenerate. Such degeneracy has been described previously in 595 G [26,27,28] and attributed to cross-relaxation due to dipole-dipole interaction. In this work we have demonstrated that the effect is general: the cross-relaxation shortens $T_1$ whenever the transition frequencies in NVs with different orientations overlap. In particular, such degeneracy takes place in zero magnetic field and is responsible for the zero-field relaxation resonance

By studying several NV⁻ samples in the [100] direction at 77 K, we have found that measurements of $T_1^{-1}$ in non-zero magnetic field can be used for estimation of the NV⁻ density (Appendix B). While the present accuracy of our estimation of the NV⁻ concentration was accurate only to about an order of magnitude, we expect that future work with samples with calibrated P1 and NV⁻ concentrations will enable a practical technique for determining local concentrations from the $T_1^{-1}$ dependences.

The presented interpretation still leaves some open question and is part of the ongoing investigation aiming at a more complete picture of the $T_1$ relaxation. On the other hand, the results presented in Fig. 7, provide confirmation that the zero-field resonance, even if overwhelmed by the phonon background for some samples becomes well visible at low temperature and for high NV⁻ density.

By using more samples prepared under different conditions, we aim at getting a more quantitative picture of the longitudinal relaxation in NV⁻ ensembles. The recently developed hole-burning methodology [34] should be useful here. The ability to select a well-defined frequency class in the inhomogeneously broadened ODMR profile with controlled application of strain and magnetic field should provide more information on various contributions to the $T_1$ relaxation.

## ACKNOWLEDGEMENTS

We thanks S. D. Chemerisov, K. Aulenbacher and V. Tyukin for irradiating the samples. This work was conducted as part of the Joint Krakow-Berkeley Atomic Physics and Photonics Laboratory and was supported by the NATO Science for Peace Programme (CBP.MD.SFP 983932), and NCN (2012/07/B/ST2/00251), MNSW (7150/E-338/M/2013), POIG (02.01.00-12-023/08 and 02.02.00-00-003/0), by the AFOSR/DARPA QuASAR program, and by the DFG DIP program.

## APPENDIX

We present here the results of our measurements of the dependence of $T_1^{-1}$ on NV$^-$ concentration outside of the region of the relaxation-rate resonances at 0 and 595 G. We extend the number of samples and the NV$^-$ concentration range over the results presented in Ref. [26]. Our aim is to determine the dependence of $T_1^{-1}$ on NV$^-$ concentration outside of the region of the relaxation-rate resonances at 0 and 595 G. The measurements were performed using the laser beam parallel to the [100] crystallographic axis. A magnetic field B = 10 G was applied along [100] to separate the $m_s=0 \rightarrow m_s=+1$ and $m_s=0 \rightarrow m_s=-1$ transitions.

We confirmed the earlier observation [26] that for each given sample there is a temperature below which $T_1^{-1}$ is independent of the temperature and confined our measurements to the range of 77-300 K. The results of the measurements (samples E1, E2, S5, and W5) are presented in Fig. A (a). The temperature dependences for these samples are compared with a phenomenological relation that takes into account the two-phonon Orbach [21] and Raman processes:

$$\frac{1}{T_1} = A_1(S) + \frac{A_2}{e^{\Delta/kT}-1} + A_3 T^5,$$

where $A_1(S)$ is a fit parameter different for each sample, $\Delta$ is the dominant local vibrational energy, and $A_2$, $A_3$, and $\Delta$ are global fit parameters common to all samples.

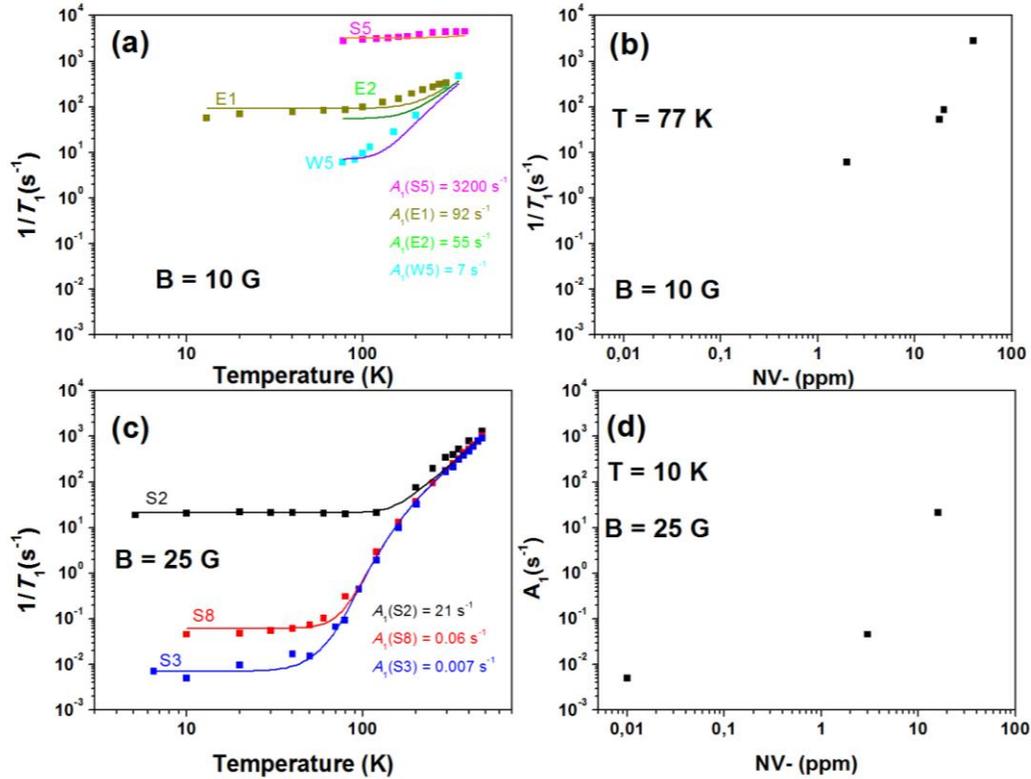

Fig. A. Relaxation rates $T_1^{-1}$ at magnetic field B=10 G directed along the [100] direction as a function of temperature (a), and as a function of the estimated NV$^-$ concentration for $T$=77 K (b). For comparison, we also plot previously reported experimental data from Ref. [26], obtained with samples S2, S3, and S8 at B=25 G for various temperatures (c), and values of $A_1$ for $T$=10 K and different NV$^-$ concentrations (d).

Figures A(b) and (d) show $T_1^{-1}$ as a function of NV$^-$ concentration. The concentration values are our best estimates, but they are only accurate to about an order of magnitude. In future work with samples with calibrated P1 and NV$^-$ concentrations; we will aim at "calibrating" the relaxation-rate dependence on the concentration, which would enable a practical technique for determining local concentrations from $T_1^{-1}$, a highly desirable and currently unavailable diagnostics for NV$^-$ ensembles.